\documentclass[conference,a4paper]{IEEEtran}
\usepackage[left=1.43cm,right=1.43cm,top=1.8cm,bottom=4.21cm]{geometry}
\IEEEoverridecommandlockouts

\usepackage[english]{babel}
\usepackage{amsmath, amssymb}
\usepackage{graphicx}
\usepackage{subfiles}
\usepackage{soul}
\usepackage{subcaption}
\usepackage{epstopdf}
\usepackage{xcolor}
\usepackage{multirow}
\usepackage{booktabs}
\usepackage{float}
\usepackage{algorithmic}

\algsetup{linenosize=\small}
\usepackage[linesnumbered,ruled]{algorithm2e}

\def\BibTeX{{\rm B\kern-.05em{\sc i\kern-.025em b}\kern-.08em
    T\kern-.1667em\lower.7ex\hbox{E}\kern-.125emX}}

\usepackage{verbatim}

\begin{document}

\title{Coordinated Multi-Armed Bandits for Improved Spatial Reuse in Wi-Fi}

\author{
\IEEEauthorblockN{Francesc Wilhelmi$^{\star\dagger}$, Boris Bellalta$^{\dagger}$, Szymon Szott$^{\mathsection}$, Katarzyna Kosek-Szott$^{\mathsection}$, Sergio Barrachina-Mu\~noz$^{\ddagger}$\vspace{0.1cm}
}
\IEEEauthorblockA{$^{\star}$\emph{Nokia Bell Labs, Stuttgart, Germany}}
\IEEEauthorblockA{$^{\dagger}$\emph{Universitat Pompeu Fabra, Barcelona, Spain}}
\IEEEauthorblockA{$^{\mathsection}$\emph{AGH University of Krakow, Poland}}
\IEEEauthorblockA{$^{\ddagger}$\emph{Centre Tecnològic de Telecomunicacions de Catalunya, Barcelona, Spain}}
\IEEEauthorblockN{\thanks{This paper is supported by the CHIST-ERA Wireless AI 2022 call MLDR project (ANR-23-CHR4-0005), partially funded by AEI and NCN under projects PCI2023-145958-2 and DEC-2023/05/Y/ST7/00004, respectively. B. Bellalta's contribution is supported by Wi-XR PID2021123995NB-I00 (MCIU/AEI/FEDER,UE) and MdMCEX2021-001195-M/ AEI /10.13039/501100011033. S. Barrachina-Muñoz is supported by Grant PID2021-126431OB-I00 funded by MCIN/AEI/ 10.13039/501100011033 (ERDF A way of making Europe) and by 6GE2E (2021 SGR 00770) from Generalitat de Catalunya.}}
}

\bstctlcite{IEEEexample:BSTcontrol}

\maketitle

\begin{abstract}
Multi-Access Point Coordination (MAPC) and Artificial Intelligence and Machine Learning (AI/ML) are expected to be key features in future Wi-Fi, such as the forthcoming IEEE 802.11bn (Wi-Fi~8) and beyond. In this paper, we explore a coordinated solution based on online learning to drive the optimization of Spatial Reuse (SR), a method that allows multiple devices to perform simultaneous transmissions by controlling interference through Packet Detect (PD) adjustment and transmit power control. In particular, we focus on a Multi-Agent Multi-Armed Bandit (MA-MAB) setting, where multiple decision-making agents concurrently configure SR parameters from coexisting networks by leveraging the MAPC framework, and study various algorithms and reward-sharing mechanisms. We evaluate different MA-MAB implementations using Komondor, a well-adopted Wi-Fi simulator, and demonstrate that AI-native SR enabled by coordinated MABs can improve the network performance over current Wi-Fi operation: mean throughput increases by 15\%, fairness is improved by increasing the minimum throughput across the network by 210\%, while the maximum access delay is kept below 3~ms.
\end{abstract}


\section{Introduction}
\label{sec:introduction}

The ever increasing demand for high performance and reliability in wireless networks has driven the development of sophisticated features for technologies like IEEE 802.11 (commercially known as Wi-Fi). This includes the ongoing work on Multi-Access Point Coordination (MAPC) for Wi-Fi~8~\cite{reshef2022future} by the IEEE 802.11bn task group, which addresses issues associated with the coexistence of multiple APs, where achieving optimal performance remains a significant challenge. MAPC entails a paradigm shift with respect to current distributed channel operations in Wi-Fi, since it is foreseen as a mechanism for enabling APs to collaborate and optimize resources across the network by exchanging information. Some potential features enabled by MAPC are Coordinated Spatial Reuse (C-SR), Coordinated Beamforming (C-BF), and Coordinated Orthogonal Frequency-Division Multiple Access (C-OFDMA)~\cite{verma2023survey}.

Regarding C-SR~\cite{wilhelmi2023throughput}, it aims to evolve the current Overlapping Basic Service Set Packet Detect (OBSS/PD) SR mechanism, which was introduced in IEEE 802.11ax (2020)~\cite{wilhelmi2021spatial}. OBSS/PD SR builds on top of BSS coloring (allowing for fast inter-BSS transmission identification) to unlock the usage of OBSS/PD thresholds that are less restrictive than Clear Channel Assessment (CCA) so that more Transmission Opportunities (TXOPs) can be created. However, OBSS/PD SR has seen little adoption in commercial equipment so far because of the moderate performance improvements compared to Distributed Coordination Function (DCF). One of the main limitations of OBSS/PD SR operation stems from its decentralized implementation, which leads to imposing too restrictive constraints in terms of transmit power, thus significantly limiting the mechanism's achievable gains. C-SR aims at overcoming OBSS/PD SR decentralization by leveraging MAPC inter-BSS communication (e.g., to exchange interference measurements), thus allowing multiple BSSs to perform efficient simultaneous transmissions. Significant research and standardization efforts are already being put into C-SR, which recently became a candidate feature for IEEE 802.11bn (Wi-Fi 8) after being accepted into the group's Specification Framework Document (SFD). However, there is still no consensus on how C-SR will materialize, so its implementation remains open.

This paper investigates the potential of MAPC-enabled coordinated Multi-Armed Bandits (MABs) for enhancing SR in coordinated Wi-Fi networks. Our approach presents a promising alternative to C-SR, with the aim of achieving similar network-wide benefits while reducing the complexity of joint Transmit Power Control (TPC) and PD adjustment decisions. MAB is a popular framework for sequential decision-making under uncertainty, which makes it well-suited for overcoming the complex OBSS interactions that occur in Wireless Local Area Networks (WLANs). This underlying complexity is precisely the main motivation for adopting Artificial Intelligence~(AI) and Machine Learning~(ML) techniques such as MABs, which are expected to address the dynamic and unpredictable nature of wireless environments, for which static approaches (even if coordinated) can fail. In fact, the adoption of AI/ML in the IEEE 802.11 is gaining momentum with the establishment of the AI/ML Study Group~(SG) and the AI/ML Standing Committee~(SC)~\cite{wilhelmi2024machine}. This paper's contributions are as follows:
\begin{itemize}
    \item We propose a coordinated MAB solution to learn the best SR policies (combining PD and transmit power configurations) on a scenario basis, so that multiple neighboring BSSs can wisely identify favorable TXOPs. Our coordinated MAB solution leverages the MAPC framework to enable inter-AP MAB communication.
    \item We propose and study a set of coordinated MAB algorithms and policies based on various reward-sharing strategies for improving SR. 
    \item We conduct simulations to study the performance of these algorithms as well as delve into their fairness. We also compare the coordinated MAB solutions to baseline approaches, including the OBSS/PD SR operation and non-coordinated bandits.
    \item We showcase the effectiveness of coordinated MAB algorithms for improving overall network performance in various Wi-Fi network scenarios and provide guidelines on how and when to use them effectively. 
\end{itemize}

\section{Related Work}
\label{sec:related_work}


The achievable performance of C-SR mechanisms has been evaluated in multiple ways and through various implementations. In~\cite{nunez2024spatial}, for example, a model-based evaluation of \mbox{C-SR} was shown to double DCF performance in some scenarios. Similarly, the work in~\cite{nunez2023multi} delved into different scheduling strategies for performing C-SR once MAPC groups are established. In that regard, the authors showed that AP-centric strategies (e.g., aiming to prioritize highly loaded APs) perform better than group-centric ones (e.g., aiming to maximize aggregate performance). Another implementation of C-SR in ns-3 demonstrated the superiority of C-SR against uncoordinated OBSS/PD SR, with a throughput improvement factor of $2.3$~\cite{imputato2024beyond}. Furthermore, a prototype solution of a centralized C-SR mechanism was presented in~\cite{haxhibeqiri2024coordinated, haxhibeqiri2024commag} to bring $33$\% goodput enhancements when enabled. Other studies confirm the appeal of applying C-SR when combined with other techniques. An example is the work in \cite{nunez2022txop}, which showed the benefits of combining C-SR with C-TDMA. In particular, C-SR with C-TDMA was shown to improve throughput by 30\% and reduce latency by a factor of 2, compared to applying C-SR only. 

Another prominent research line, as part of future AI-native radios~\cite{wilhelmi2024machine}, is AI-driven SR, which aims to overcome the limitations of the static approaches currently proposed for C-SR. In particular, existing proposals use radio measurements such as the Received Signal Strength Indicator (RSSI) statically (e.g., averaged values) to derive C-SR policies that allow for an acceptable Signal-to-Noise Ratio (SNR). However, such an approach may lose effectiveness in real deployments where the variability of both RSSI and SNR is very high. Some works exploring AI-driven SR solutions are \cite{wilhelmi2019potential, wilhelmi2019collaborative}, where different MAB solutions were proposed and studied to drive the optimization of SR parameters such as PD and transmit power. The exploration-exploitation paradigm embedded in MAB showed potential for addressing the non-stationarity experienced in OBSSs implementing SR solutions. However, these solutions were based on decentralized mechanisms (no communication between agents was provided), which led to several issues as a result of the competition arising among networks. MAB solutions have also been assessed in other Wi-Fi problems, such as channel bonding, demonstrating their effectiveness as a lightweight, efficient, ready-to-use solution~\cite{barrachina2021multi}. In this work, similar to what was done in~\cite{wilhelmi2019potential, wilhelmi2019collaborative}, we focus on an AI-native solution to improve SR. As a step forward, we study the advantages that coordination may bring to the operation of decentralized agents, which are assumed to cooperate within an MAPC framework.

\section{Coordinated MAB}
\label{sec:solution}

We consider a Multi-Agent MAB (MA-MAB) setting in which multiple agents are individually enrolled in different BSSs for optimizing SR by tuning the PD (which determines to which extent inter-BSS interference can be ignored) and the transmit power (which determines to which extent inter-BSS interference is created). The proposed solution departs from the operations defined for OBSS/PD-based SR, which was detailed in~\cite{wilhelmi2021spatial}, hence making it compatible with currently available products. As a step further, and by leveraging MAPC capabilities, we allow agents to exchange feedback about their performance regularly, thus aiming at creating collaborative strategies that enhance global performance. Formally, based on some action--selection strategy $\mathcal{E}$ (cf. Section~\ref{sec:action_selection}), each player (or agent) $p\in \mathcal{P}$ sequentially plays an action (or arm) $k\in \mathcal{A}^{(p)}$ at each time step $t=1,...,T$ and obtains a reward $r^{(p)}(t+1) \in \mathbb{R}$ that depends on the joint action profile $K(t)= \{k^{(1)}(t), ..., k^{(|\mathcal{P}|)}(t)\}$ and on the reward calculation strategy $\mathcal{R}$ (cf. Section~\ref{sec:reward_calculation}). For the SR problem, we consider actions composed of discrete PD ($\gamma$) and transmit power ($\zeta$) values, which leads to an individual action space $\mathcal{A}^{(p)}_{|\gamma|\times|\zeta|}$ and a global action space $\mathcal{A}_{|\mathcal{P}|\times|\gamma|\times |\zeta|}$.

\begin{figure}[ht!]
    \centering
    \includegraphics[width=0.8\columnwidth]{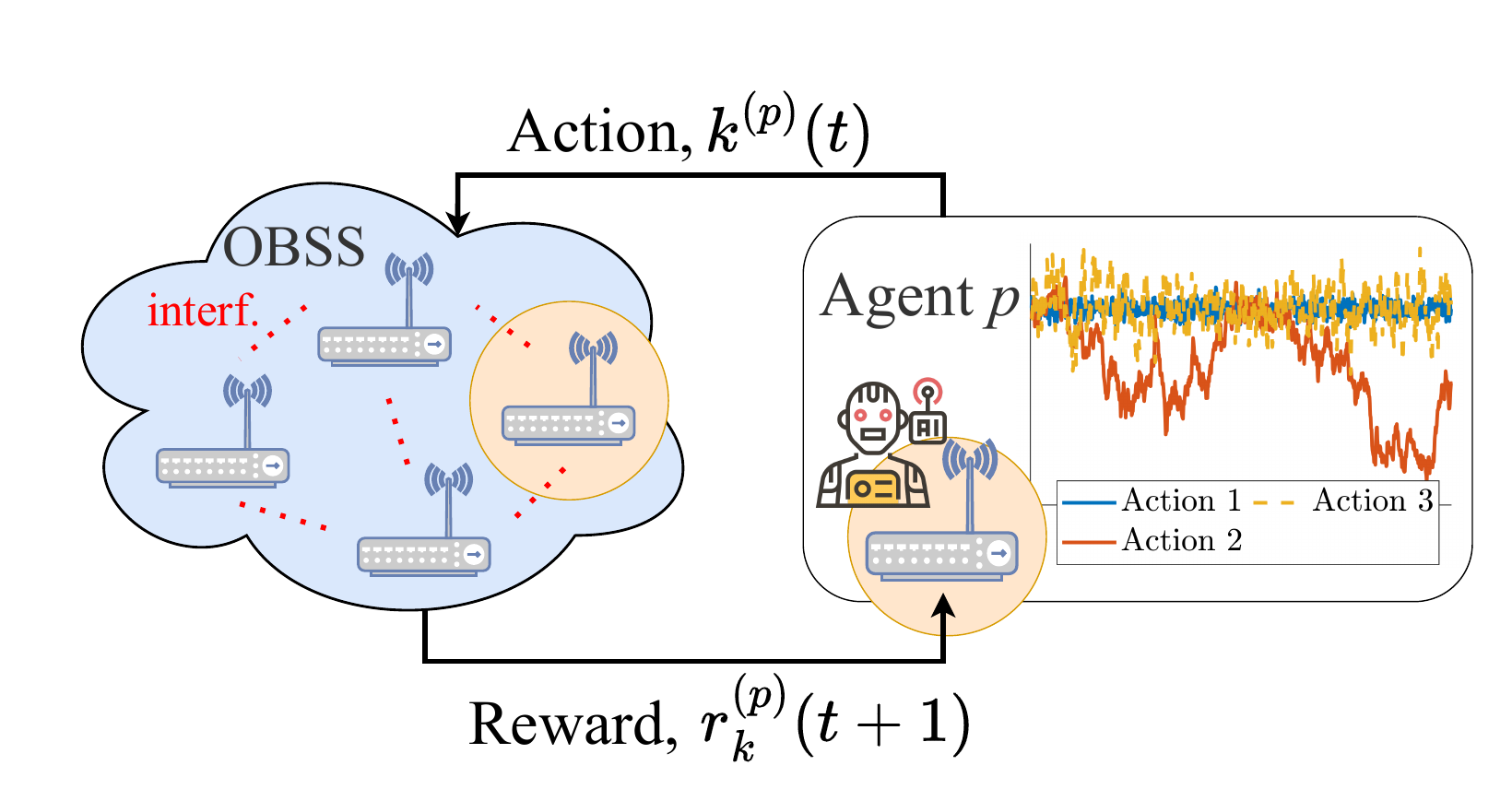}
    \caption{Representation of an agent's operation in an OBSS, where the rewards associated with the different available actions are drawn from an unknown distribution.}
    \label{fig:agents_operation}
\end{figure}

MAB operation is illustrated in Figure~\ref{fig:agents_operation} and is generically described in Algorithm~\ref{alg:bandit}. After initializing the rewards $r_k^{(p)}$ and the number of plays $N_k^{(p)}$ for each action $k\in \mathcal{A}^{(p)}$, agent $p$ sequentially selects an action based on the action-selection strategy $\mathcal{E}$ (line~4 in Algorithm~\ref{alg:bandit}) and, in the next iteration, receives a reward based on the reward calculation strategy $\mathcal{R}$ (line~3 in Algorithm~\ref{alg:bandit}). The simultaneous operation of multiple agents breaks the stationarity of the rewards' distributions, thus making the problem challenging. Nevertheless, the multi-agent setting keeps the complexity of the problem low, which otherwise would lead to a combinatorial action space.

\begin{algorithm}[h!]
	\SetKwInOut{Input}{Input}
	\SetKwInOut{Output}{Output}		
	\textbf{Initialize:} $t=0$, for each arm $k \in \mathcal{A}^{(p)}$, set $r^{(p)}_{k} = 0$ and $N^{(p)}_k = 0$ \\
	\For{$t = 1...T$}
	{
         Update the reward $r_{k}(t)$ based on the performance observed in $t+1$ and $\mathcal{R}$ \\
         Select arm $k\in \mathcal{A}^{(p)}$ based on $\mathcal{E}$\\
         $N^{(p)}_{k} \leftarrow N^{(p)}_{k} + 1$\\
	}
	\caption{MAB implementation by agent $p$.}
	\label{alg:bandit}
\end{algorithm}	

\subsection{Action Selection Strategy ($\mathcal{E}$)}
\label{sec:action_selection}

We focus on two exploration--exploitation strategies, namely $\varepsilon$-greedy~\cite{auer2002finite} and Thompson sampling~\cite{thompson1933likelihood}, for driving the training of agents towards finding the best actions. First, $\varepsilon$-greedy is a classic strategy used in reinforcement learning whereby an agent samples an arm uniformly at random ($\mathcal{U}$) with a certain fixed probability (exploration) and selects the arm yielding the best-observed performance the rest of the time (exploitation). The $\varepsilon$-greedy algorithm, despite its simplicity, has been shown to be effective in complex settings such as OBSS~\cite{barrachina2021multi}. In particular, an agent implementing $\varepsilon$-greedy selects an arm as:
\begin{equation}
    k=\begin{cases}      
    k \sim \mathcal{U}(1, |\mathcal{A}^{(p)}|), & \text{with prob. } \varepsilon~ \text{(exploration)}, \\ 
    \underset{k\in\mathcal{A}^{(p)}}{\text{argmax }} r^{(p)}_{k}, & \text{with prob. } 1 - \varepsilon~\text{(exploitation)}.
    \end{cases}
\end{equation}

Thompson sampling addresses exploration-exploitation in a different way than $\varepsilon$-greedy: that it samples arms based on their probability of being optimal, according to a probabilistic model constructed from the observed rewards. In particular, as previously done in \cite{wilhelmi2019potential}, we assume that the distribution of the reward associated with each arm is given by a Gaussian distribution $\mathcal{N}$. Based on that, an agent implementing Thompson sampling selects, in each iteration, the arm maximizing $\text{argmax }_{k\in\mathcal{A}^{(p)}} \theta_k$, where
\begin{equation}
    \theta_{k\in\mathcal{A}^{(p)}}\sim \mathcal{N}\bigg(\hat{r}_{k}^{(p)}, \frac{1}{N_k^{(p)} + 1}\bigg),
\end{equation}
where the estimated reward $\hat{r}_{k}^{(p)}$ is given by
\begin{equation}
    \hat{r}_{k}^{(p)} \leftarrow \frac{\hat{r}_{k}^{(p)}  N^{(p)}_{k} + r^{(p)}_{k}}{r^{(p)}_{k} + 2}.
\end{equation}

\subsection{Reward Calculation Strategy ($\mathcal{R}$)}
\label{sec:reward_calculation}

To compute the reward, we consider different sharing strategies, which are based on a selfish reward (\texttt{SELF}) that is firstly computed by each agent individually. In particular, the selfish reward is calculated as the normalized individual throughput $r^{(p)}=\Gamma^{(p)}/\Gamma^{*(p)}$, where $\Gamma^{*(p)}$ is the maximum achievable throughput in isolation, derived from the maximum transmission capabilities of the BSS given by the best applicable Modulation and Coding Scheme (MCS).\footnote{Other performance indicators such as latency can be adopted as a reward.} As for the sharing strategies, enabled by the MAPC framework, they allow the different agents to play actions according to a shared performance among the players. From the MAB algorithms' perspective, MAPC is assumed to allow for a perfect monitoring setting, which means that every agent has access to the rewards experienced by others. According to this, we study the following shared reward criteria:
\begin{itemize}
    \item \textbf{Average, \texttt{AVG}:} The shared reward is calculated as the average value of each individual reward, so that $r_\texttt{AVG} = \frac{1}{|\mathcal{P}|}\sum_{p=1}^{|\mathcal{P}|} r^{(p)}$.
    \item \textbf{Max-min, \texttt{MAX-MIN}:} The shared reward is computed as the minimum value of each individual reward, so that $r_\texttt{MAX-MIN} = \min_{p\in \mathcal{P}} (r^{(p)})$.
    \item \textbf{Proportional fairness, \texttt{PF}:} The shared reward is calculated as the sum of the logarithms of each individual reward, so that $r_\texttt{PF} = \sum_{p=1}^{|\mathcal{P}|} \log(r^{(p)})$.
\end{itemize}

In this work, we assume that the communication between agents for sharing the rewards is negligible and lossless, and occurs at the end of each learning iteration of fixed duration $\Delta$. Future work is expected to shed light on the overheads associated with the coordinated MABs, as well as on the implications of sharing information through wireless links exposed to contention and interference effects.

\section{Performance Evaluation}
\label{sec:performance_evaluation}

In this section, we study the different strategies for improving SR through simulations. We use Komondor~\cite{barrachina2019komondor}, an IEEE 802.11 simulator with embedded AI-capable agents. The various MAB implementations were simulated along with the network operation using the parameters collected in Table~\ref{tbl:simulation_parameters}. More details on the considered IEEE 802.11 frame types, sizes, and inter-frame periods are in~\cite[Table~B.6]{wilhelmi2021spatial}.

\begin{table}[ht!]
\centering
\caption{Simulation parameters.}
\label{tbl:simulation_parameters}
\resizebox{.92\columnwidth}{!}{%
\begin{tabular}{@{}clc@{}}
\toprule
Parameter & \multicolumn{1}{c}{\textbf{Description}} & \textbf{Value} \\ \midrule
$t$ & Simulation time & $300$ s \\
$F_c$ & Carrier frequency & $5$ GHz \\
$\text{GI}$ & Guard Interval & $3.2$ $\mu$s\\ 
$B$ & Transmission bandwidth & $20$ MHz \\
$\text{MCS}$ & MCS indices & 0-11\\ 
$\mathcal{P}^\text{Noise}$ & Noise power & $-95$ dBm \\
$\mathcal{P}_{tx,\max}$ & Default transmit power & $20$ dBm \\
CCA & Default CCA threshold & $-82$ dBm \\
$S$ & Single-user spatial streams & $1$ \\
$G^\text{TX/RX}$ & Transmitter/receiver antenna gain & $0/0$ dBi \\
$\textrm{CE}$ & Capture effect threshold & $10$ dB \\
PL & Path loss model & See \cite{wilhelmi2021spatial} \\
$PL_0$ & Loss at the reference dist. & $5$ dB \\
$\nu$ & Path-loss exponent & $4.4$ \\
$\sigma$ & Shadowing factor & $9.5$ dB\\
$\omega$ & Obstacles factor & $30$ dB\\
$\text{TXOP}_{\max}$ & TXOP duration limit & $5.484$ ms\\ 
$\text{A-MPDU}_{\max}$ & A-MPDU size & $64$\\ 
$L_{D}$ & Length of data packets & $1500$ bytes \\ 
$\mathcal{T}$ & Traffic model & Full-buffer\\ 
$\mathbb{T}$ & Traffic type & Downlink (DL) \\ 
$\text{CW}_0$ & Initial Contention Window (CW) & $16$ \\
$\text{CWE}_{\min/\max}$ & Min./Max. CW exponent & $1/5$ \\
\midrule
$\gamma$ & PD values & $\{-72, -82\}$~dBm\\
$\zeta$ & Transmit power values & $\{10, 20\}$~dBm\\
$\Delta$ & Iteration duration & $0.5$~s \\
$\varepsilon_0$ & Initial exploration coefficient & $0.1$ \\
$\varepsilon(t)$ & Exploration coefficient & $\varepsilon_0/\sqrt{t}$ \\
\bottomrule
\end{tabular}%
}
\end{table}

\subsection{Interactions between 2 BSSs}

\begin{figure}[ht!]
    \centering
    \includegraphics[width=\linewidth]{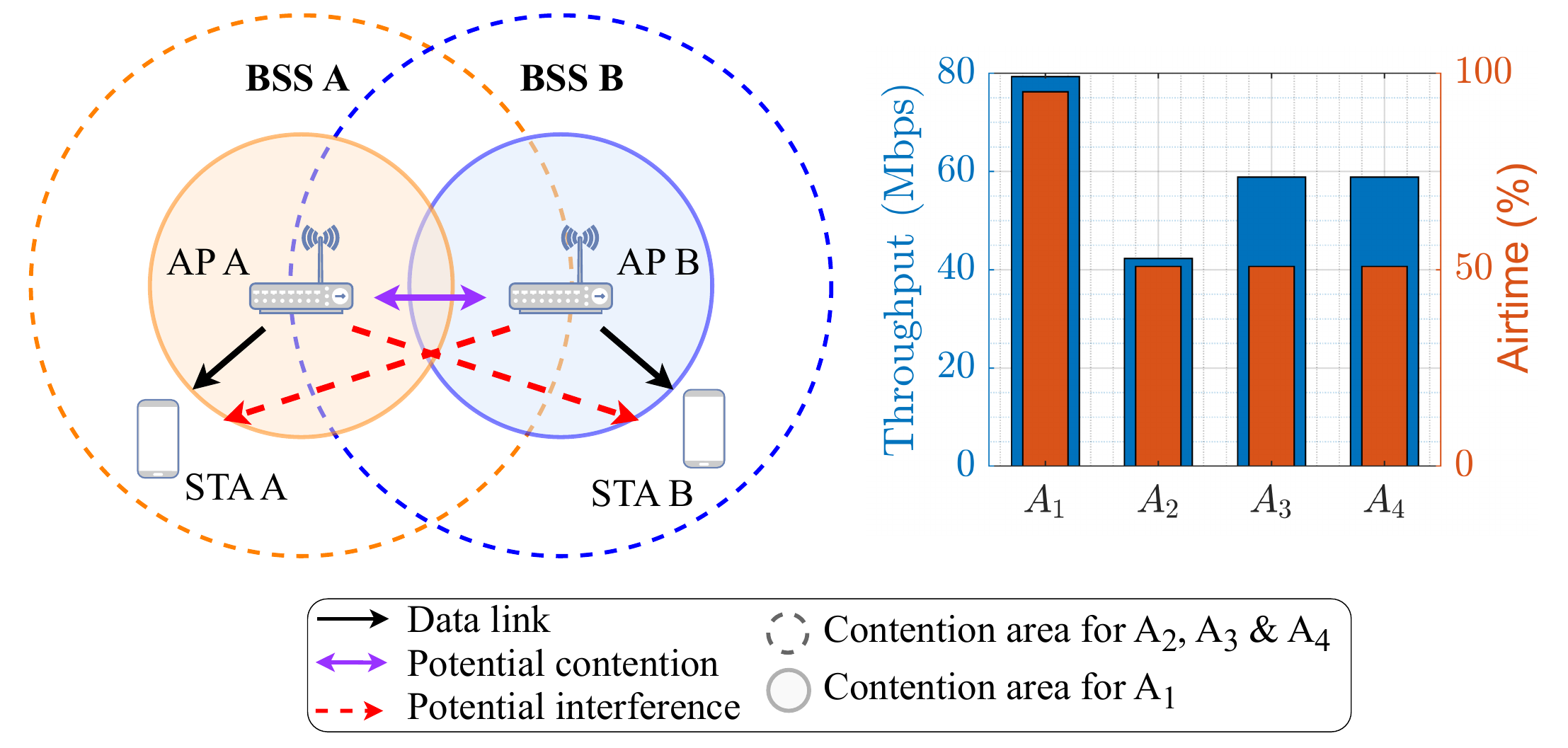}
    \caption{Considered 2-BSS toy deployment and mean performance of $A_1=\{10,-72\}$~dBm, $A_2=\{10,-82\}$~dBm, $A_3=\{20,-72\}$~dBm, $A_4=\{20,-82\}$~dBm.}
    \label{fig:toy_scenario}
\end{figure}

We start with the deployment depicted in Figure~\ref{fig:toy_scenario} (later referred to as the \textit{toy scenario}), which contains two BSSs where different contention and interference interactions occur depending on the power and PD used by the BSSs. To keep complexity low and clearly devise the effect of applying different bandit strategies, we consider four possible configurations (or actions) combining  $\gamma = \{-72, -82\}$~dBm and $\zeta = \{10,20\}$~dBm, thus leading to the per-agent action space $\mathcal{A} = \gamma \times \zeta$. Figure~\ref{fig:toy_scenario} also shows the performance in terms of throughput and airtime in the particular case where the two BSSs jointly apply each of the $|\mathcal{A}|=4$ considered configurations. Given the symmetry of the considered deployment, the optimal performance is obtained when the two BSSs use configuration $\{k^{(1)}, k^{(2)}\} = \{A_1, A_1\}$.

\begin{figure}[ht!]
\centering
\includegraphics[width=\columnwidth]{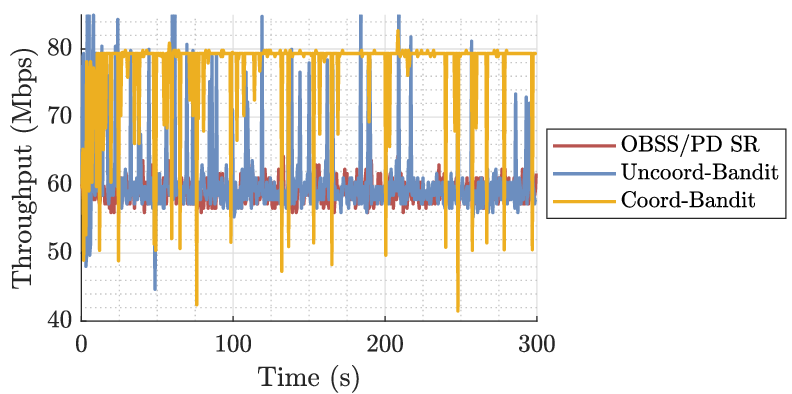}
\caption{Average throughput observed in the toy scenario for OBSS/PD SR, uncoordinated bandit with $\varepsilon$-greedy, and coordinated bandit with $\varepsilon$-greedy (\texttt{AVG}).}
\label{fig:bandits_performance_toy_scenario}
\end{figure}

Next, we show the performance achieved by the BSSs of the toy scenario when using coordinated bandits with $\varepsilon-$greedy as an action-selection strategy $\mathcal{E}$ and \texttt{AVG} as the sharing reward policy $\mathcal{R}$. Figure~\ref{fig:bandits_performance_toy_scenario} shows the mean throughput achieved during the simulation by the coordinated bandit solution and two baselines, OBSS/PD SR with $\text{OBSS/PD}_{thr} = -72$ dBm, and uncoordinated bandits using $\varepsilon$-greedy and $\texttt{SELF}$. In addition, Table~\ref{tab:selected_actions_toy_scenario} contains the percentage of times each action was selected by each mechanism during the simulation.

\begin{table*}[ht!]
\centering
\caption{Selected actions by uncoordinated (Uncoord-Bandit) and coordinated bandits (Coord-Bandit) in the toy scenario.}
\label{tab:selected_actions_toy_scenario}
\resizebox{.7\textwidth}{!}{%
\begin{tabular}{@{}cccccccc@{}}
\toprule
\multirow{2}{*}{} & \multirow{2}{*}{\textbf{\begin{tabular}[c]{@{}c@{}}Uncoord-Bandit\\ ($\varepsilon$-greedy)\end{tabular}}} & \multicolumn{3}{c}{\textbf{Coord-Bandit ($\varepsilon$-greedy)}} & \multicolumn{3}{c}{\textbf{Coord-Bandit (Thompson sampling)}} \\ \cmidrule(l){3-8} 
 &  & \texttt{AVG} & \texttt{MAX-MIN} & \texttt{PF} & \texttt{AVG} & \texttt{MAX-MIN} & \texttt{PF} \\ \midrule
\textbf{$A_1$} & 4.51\% & 92.49\% & 2\% & 92.48\% & 60.47\% & 58.43\% & 58.43\% \\
\textbf{$A_2$} & 1.67\% & 2.42\% & 2.33\% & 2.75\% & 8\% & 10.85\% & 10.85\% \\
\textbf{$A_3$} & 1.67\% & 2.34\% & 5.75\% & 2.33\% & 8\% & 6.59\% & 6.59\% \\
\textbf{$A_4$} & 92.15\% & 2.75\% & 89.89\% & 2.42\% & 27.58\% & 24.12\% & 24.12\% \\ \bottomrule
\end{tabular}%
}
\end{table*}

As shown in Figure~\ref{fig:bandits_performance_toy_scenario}, the coordinated MABs allow for maximizing the average performance when compared to OBSS/PD SR and uncoordinated bandits. This is because the coordinated mechanism drives the two BSSs playing $A_1$ most of the time (see Table~\ref{tab:selected_actions_toy_scenario}), which matches with the optimal configuration shown in Figure~\ref{fig:toy_scenario}. In contrast, OBSS PD SR statically uses the same configuration and the uncoordinated MABs get stuck in the most conservative action ($A_4$), as a result of the competition among the two BSSs. This effect was previously shown in~\cite{wilhelmi2019collaborative} and occurs when optimal actions lead to poor performance when the other agent selects a different action. As a result, agents using the \texttt{SELF} reward get stuck in a weak equilibrium by conservatively selecting $A_4$. This result motivates the adoption of cooperative strategies.

\begin{figure}[ht!]
\centering
\includegraphics[width=.75\columnwidth]{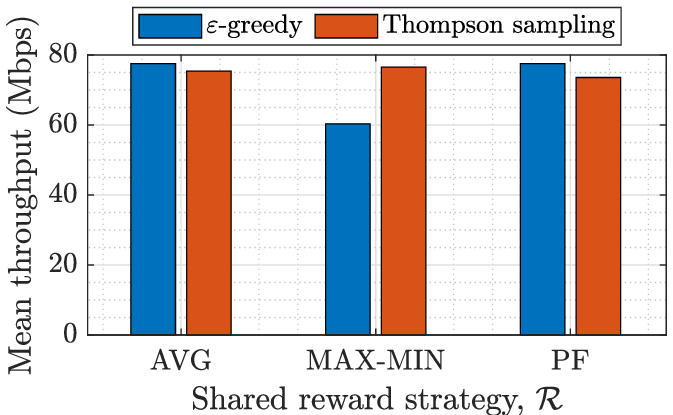}
\caption{Mean average throughput observed in the toy scenario for $\mathcal{E} = \{\varepsilon\text{-greedy}, \text{Thompson sampling}\}$ and $\mathcal{R} = \{\texttt{AVG}, \texttt{MAX-MIN}, \texttt{PF}\}$.}
\label{fig:comparison_of_policies}
\end{figure}

\begin{figure}[ht!]
\centering
\includegraphics[width=\columnwidth]{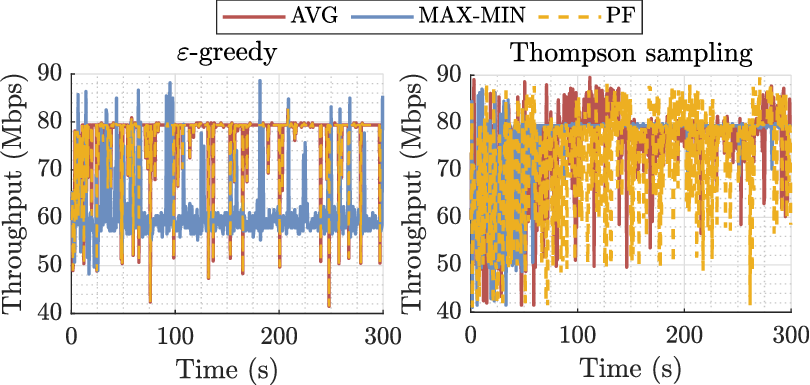}
\caption{Average throughput observed in the toy scenario when applying $\varepsilon$-greedy (left) and Thompson sampling (right), for $\mathcal{R} = \{\texttt{AVG}, \texttt{MAX-MIN}, \texttt{PF}\}$.}
\label{fig:comparison_of_policies_temporary_throughput}
\end{figure}

Still considering the toy scenario (Figure~\ref{fig:toy_scenario}), we study the performance achieved when using different action-selection strategies $\mathcal{E}$ ($\varepsilon$-greedy and Thompson sampling) and shared rewards $\mathcal{R}$ (\texttt{AVG}, \texttt{MAX-MIN}, and \texttt{PF}). For each combination, we show the mean average throughput (including the performance experienced in the transitory phase) of the two BSSs (Figure~\ref{fig:comparison_of_policies}), and the mean throughput evolution (Figure~\ref{fig:comparison_of_policies_temporary_throughput}).

As shown in Figure~\ref{fig:comparison_of_policies}, $\varepsilon$-greedy leads to better performance than Thompson sampling when \texttt{AVG} and \texttt{PF} are used. Figure~\ref{fig:comparison_of_policies_temporary_throughput} shows that this is because Thompson sampling explores much more intensively than the considered $\varepsilon$-greedy implementation. Under \texttt{MAX-MIN}, $\varepsilon$-greedy reaches the weak equilibrium mentioned above because \texttt{MAX-MIN} excessively penalizes unfair global configurations, even if those include the best action ($A_1$) by one of the agents. Thompson sampling, in contrast, overcomes this issue thanks to the probabilistic reward models built on top of each action, which allows selecting $A_1$ more times than $\varepsilon$-greedy. The previous observations are reinforced by the results in Table~\ref{tab:selected_actions_toy_scenario}, where $A_1$ ends up being the most played action by Thompson sampling, even for \texttt{MAX-MIN}. However, unlike for $\varepsilon$-greedy (\texttt{AVG} and \texttt{PF}), $A_4$ is also significantly played by Thompson sampling during the transitory phase, i.e., during the first $100$~seconds of the simulation, thus affecting the overall performance in that period.

\subsection{Random scenarios}

\begin{figure}[ht!]
    \centering
    \includegraphics[width=\linewidth]{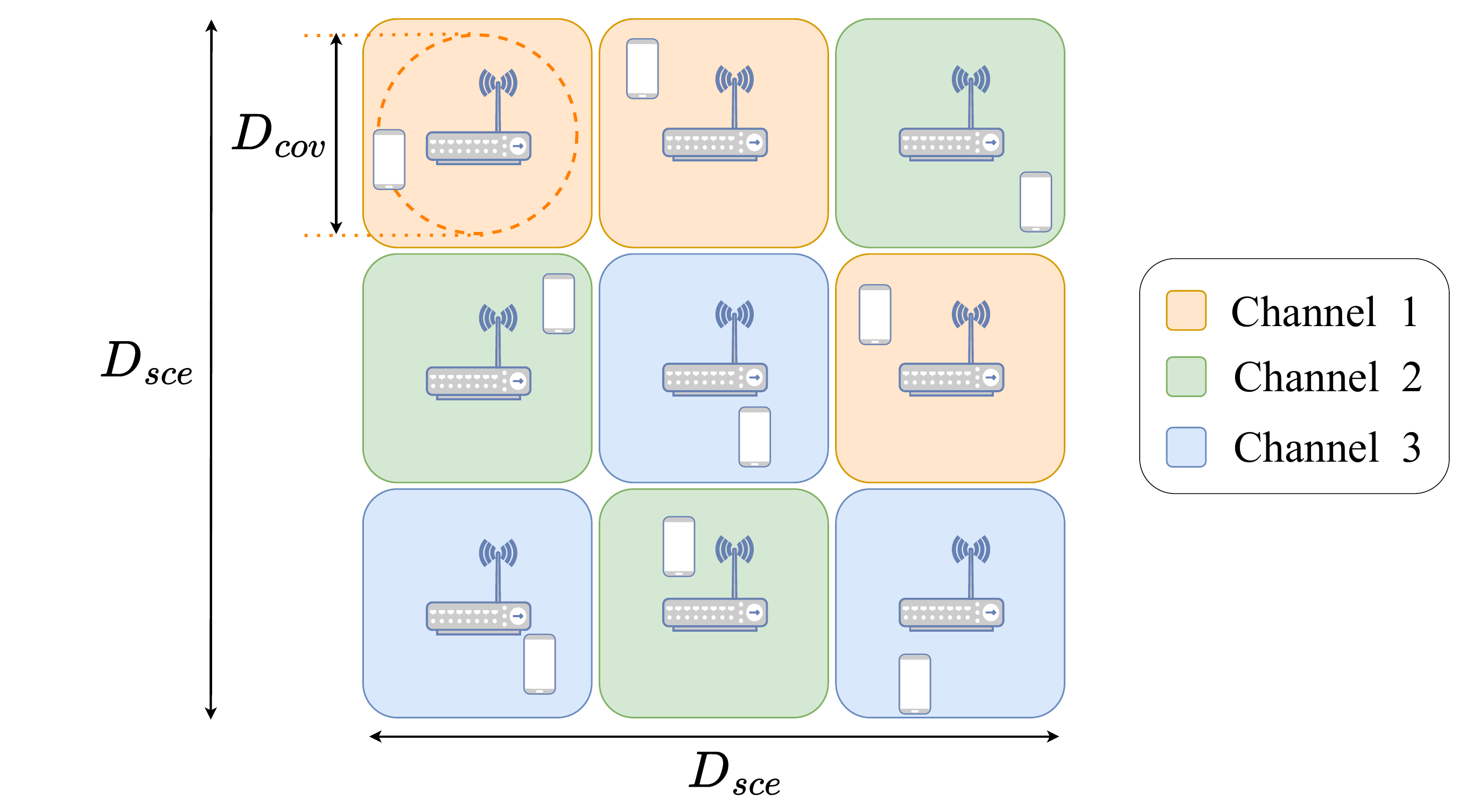}
    \caption{Considered 9-BSS random deployment.}
    \label{fig:randm_scenario}
\end{figure}

Next, we provide more insights into the different considered cooperative strategies by simulating the performance of various BSSs at larger and denser random deployments. For that, we consider the scenario shown in Figure~\ref{fig:randm_scenario}, where $9$~BSSs are deployed in a $3\times 3$ grid of side $D_{sce}=20$~m. Each AP is deployed deterministically in the center of each cubicle and their associated STAs ($1$ per AP) are randomly deployed around their APs within a coverage area with diameter $D_{cov}=3$~m. A frequency reuse $F_R = 3$ means that each BSS is allocated with a random non-overlapping 20~MHz channel from $\mathcal{C} = \{1,2,3\}$. $R=100$ random drops are considered.

\begin{figure*}[ht!]
    \centering
    \includegraphics[width=.95\linewidth]{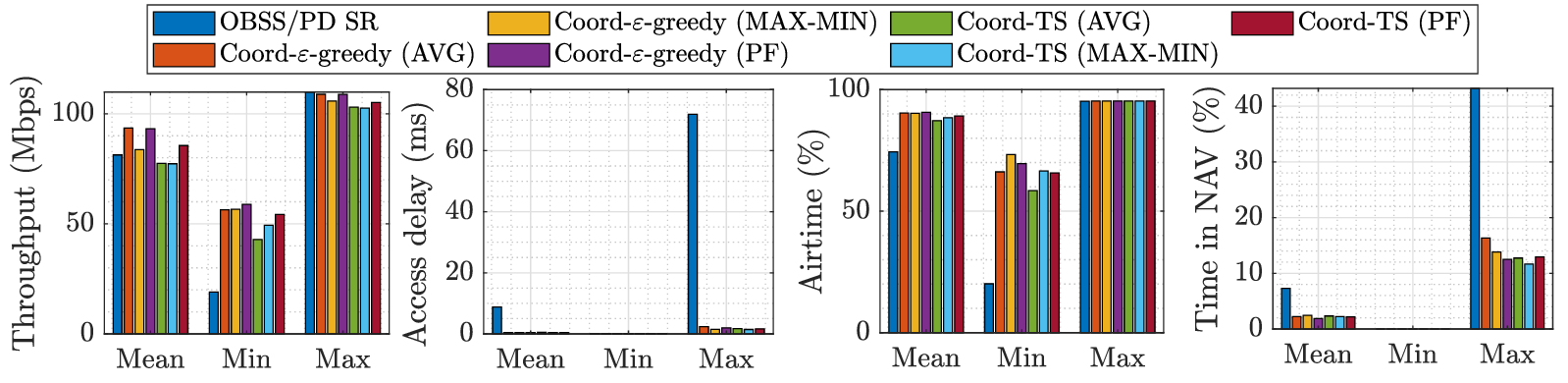}
    \caption{Mean performance achieved in the $9$-BSS deployment by $\mathcal{E} = \{\varepsilon\text{-greedy}, \text{Thompson sampling}\}$ and $\mathcal{R} = \{\texttt{AVG}, \texttt{MAX-MIN}, \texttt{PF}\}$. Left: throughput, left-center: average delay, right-center: airtime, right: NAV time.}    \label{fig:bar_plot_mean_performance_all}
\end{figure*}

Figure~\ref{fig:bar_plot_mean_performance_all} shows the mean performance achieved by the OBSS/PD SR baseline and each coordinated algorithm using the different shared reward strategies. In particular, we show the mean values obtained across all devices in the scenario throughout all the simulated random drops, in terms of throughput, access delay, airtime, and time spent in virtual carrier sensing, referred to as the Network Allocation Vector (NAV) time. In all the cases, we display the mean, minimum, and maximum performance experienced across the $9$ BSSs of each simulated deployment.

Starting with throughput, we observe that the coordinated MAB implementations slightly improve the mean performance (up to $15$\% for Coord-$\varepsilon$-greedy-\texttt{AVG}) but lead to slightly lower maximum performance (up to $-6$\% for Coord-TS-\texttt{MAX-MIN}). However, the minimum performance of OBSS/PD SR is significantly outperformed by all MAB implementations (up to $210$\% for Coord-$\varepsilon$-greedy-$\texttt{PF}$), which evidences the lack of fairness experienced by the OBSS/PD SR (some BSSs starve) and the potential of coordinated MAB to address this. The same effects are noticed for the rest of the performance metrics. In particular, we notice a great advantage in terms of maximum access delay, which for OBSS/PD SR reaches, on average, $58$~ms, while for the different MAB implementations, it is kept under $3$~ms. This is translated into the minimum airtime experienced by starved devices, which with coordinated MAB enjoy up to $73$\% airtime. Another important metric is the NAV time, which prevents devices from accessing the medium when receiving Ready-to-Send (RTS) and Clear-to-Send (CTS) frames from other overlapping BSSs. Inversely to the airtime, the maximum NAV time is substantially decreased by the coordinated MABs when compared to the OBSS/PD SR. Finally, with respect to the different policies adopted, we observe a similar trend as in the previously described toy scenario. When using $\varepsilon$-greedy, the best-performing reward-sharing strategies are \texttt{AVG} and \texttt{PF}. Nevertheless, \texttt{MAX-MIN} is useful to improve the minimum throughput across the OBSS. As for Thompson sampling, the behavior is slightly different, due to its different approach to handling exploration. In that case, \texttt{PF} is shown to be the best-performing option.

\begin{figure}[t!]
    \centering
    \includegraphics[width=\linewidth]{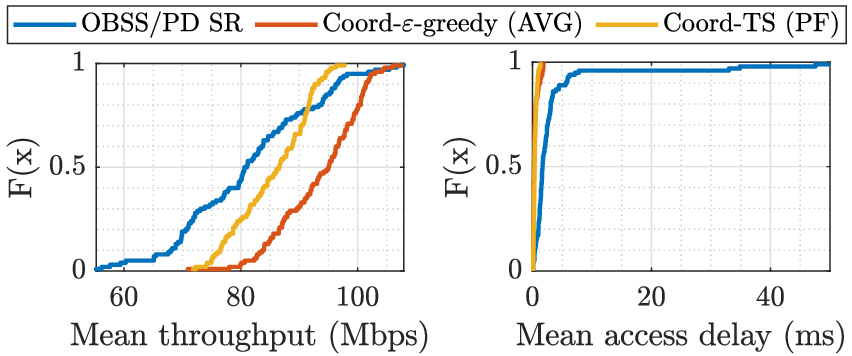}
    \caption{CDF of the mean throughput (left) and the access delay (right) achieved in the $9$-BSS deployment by OBSS/PD SR, Coord-$\varepsilon$-greedy (\texttt{AVG}) and Coord-TS (\texttt{PF}).}
    \label{fig:cdf_mean_throughput_best}
\end{figure}

Based on previous observations, we focus on the performance achieved by the best-performing options for coordinated bandits, namely $\varepsilon$-greedy (\texttt{AVG}) and Thompson sampling (\texttt{PF}). Figure~\ref{fig:cdf_mean_throughput_best} shows the Cumulative Distribution Function (CDF) of the mean throughput and channel access delay experienced throughout all the simulations. The two MAB implementations grant higher throughput than OBSS/PD SR,  $\varepsilon$-greedy (\texttt{AVG}) being the best one (Figure~\ref{fig:cdf_mean_throughput_best}). In some cases, OBSS/PD SR leads to slightly higher throughput than Thompson sampling (\texttt{PF}), but this stems from the unfair situations observed above. For mean access delay, the two MAB implementations lead to similar results, but in both cases OBSS/PD SR is substantially outperformed.

\section{Conclusions}
\label{sec:conclusions}

This paper explored the potential of coordinated MA-MAB to improve SR in MAPC-equipped Wi-Fi networks. Our findings demonstrate that coordinated MABs offer a significant advantage over the OBSS/PD SR operation and uncoordinated MABs. We studied different exploration-exploitation strategies and showed that explicit exploration ($\varepsilon$-greedy) grants control on the exploration but might not adapt properly to different situations, whereas implicit exploration (Thompson sampling) leads to higher instability but adapts well to various contexts and reward types, eventually leading to optimal policies. In addition, we studied different reward-sharing strategies and showed that metrics such as \texttt{AVG} and \texttt{PF} can drive agents to find global configurations that maximize overall performance. Meanwhile \texttt{MAX-MIN} is useful to maximize minimum performance, but fails to maximize overall performance in some situations. Future work will study the feasibility of the proposed solution through a detailed analysis of overheads and computations associated with the bandits' operation. 

\bibliographystyle{IEEEtran}
\bibliography{bib}

\end{document}